\documentclass[11pt, onecolumn]{IEEEtran}

\usepackage{graphicx,amsmath,amssymb,cite}

\usepackage[dvips]{color}
\usepackage{float}
\ifCLASSINFOpdf
\else
\fi
\begin{document}
\title{Probability density derivation and analysis of SINR in massive MIMO systems with MF beamformer}

\author{Shu Feng, Gu Chen,  Wang Mao, Stevan Berber, and You Xiaohu

\thanks{This work was supported in part by open research fund of National Mobile Communications Research Laboratory, Southeast University (Grant No. 2013D02), the Fundamental Research Funds for the Central Universities (Grant No. 30920130122004), and the National Natural Science Foundation of China (Grants No. 61271230, and No. 61472190).}
\thanks{Shu Feng, Gu Chen, and Wang Mao are with the School of Electronic and Optical Engineering, Nanjing University of Science and Technology, Nanjing, China. Stevan Berber is with the University of Auckland in New Zealand. You Xiaohu, Wang Mao, and Shu Feng are  with the National Mobile Communications Research Laboratory, Southeast University. Shu Feng is also with the Ministerial Key Laboratory of JGMT, NJUST.(email: shufeng@njust.edu.cn).}
\thanks{}
}

%



\maketitle

\begin{abstract}
In massive MIMO systems, the matched filter (MF) beamforming is attractive technique due to its extremely low complexity of implementation compared to those high-complexity decomposition-based beamforming techniques such as zero-forcing,  and  minimum mean square error. A specific problem in applying these techniques is how to qualify and quantify the relationship between the transmitted signal, channel noise and interference. This paper presents  detailed procedure of deriving an approximate formula for probability density function (PDF) of the signal-to-interference-and-noise ratio (SINR) at user terminal when multiple antennas and  MF beamformer are used at the base station. It is shown how the derived density function of SINR can be used to calculate the symbol error rate of massive MIMO downlink. It is confirmed by simulation that the derived approximate  expression for PDF is consistent with the simulated PDF in medium-scale and large-scale MIMO systems.
\end{abstract}




\IEEEpeerreviewmaketitle

\newpage
\section{Introduction}
\IEEEPARstart{R}{ecently}, massive multi-user MIMO (MU-MIMO) is an up-to-date active research area in wireless communications  due to its extremely high multiplexing and diversity gains, and fine spatial resolution \cite{rusek,marz,larsson,jacob}. It has been viewed as a strong candidate for key techniques of the future-generation wireless networks. In traditional MU-MIMO scenarios, several classic linear precoding schemes are developed like matched filter (MF), zero-forcing (ZF)\cite{spencer, wiesel}, block diagonalization (BD)\cite{choi}, maximum signal-to-leakage-plus-noise ratio (Max-SLNR)\cite{sadek}, singular value decomposition (SVD)\cite{liu}, and channel inversion plus regularization including minimum mean square error (MMSE)\cite{peel, hak}. These precoders can be readily extended to  massive MU-MIMO systems. However, the beamforming performance in massive MU-MIMO is intimately related to pilot structures and degree of precision of channel estimation \cite{jacob,noh,mar1}. In \cite{noh}, for the assumed channel characterized by the Gauss-Markov random process, a novel pilot design criterion is proposed to minimize the channel estimation mean square error such that sum-rate (channel capacity) is improved. In \cite{mar1}, an orthogonal pilot transmit scheme is proposed for multi-cell environment. Compared to non-orthogonal pattern, it can achieve an over 15dB signal-to-interference-and-noise ratio (SINR) gain.


The decomposition-based precoders such as ZF, BD, MMSE and Max-SLNR require about $KN^3$ complex multiplications (CMs) where $N$ and $K$ stand for the number of antennas at base station (BS) and the number of users in system, respectively. Thus, precoding complexity is a challenging issue as the number of antennas at BS approaches several hundred even one thousand. In \cite{park}, authors design a reduced-complexity ZF precoder for massive MIMO. A truncated polynomial expansion of matrix inversion in \cite{kamm} is applied to the regularized zero forcing to reduce the computational amount of matrix inversion and the corresponding coefficients are optimized to maximize the SINR in multi-cell MIMO system. However, the complexity of these precoding algorithms is still much larger than $KN$ CMs (Complexity of MF). As a low-complexity precoder, MF tends to the optimal linear precoder SVD in the sum-rate sense from \cite{rusek,marz,larsson,jacob} as $N$ approaches large-scale.

In the development of precoding schemes the main problem is how to find the probability density function of SINR in a massive MIMO system with MF precoder at  BS. The solution of this problem is very important,because this PDF is necessary to compute  the probability of symbol error and outage performance of massive MIMO. For this reason,  an approximate expression of the PDF formula for SINR is derived and confirmed by simulation.

This paper is organized as follows. System model is described in  Section II. Section III presents the detailed deriving process and skills of the PDF of SINR. The numerical results are shown and discussed in section IV, and Section V concludes this paper.

Notations: throughout the paper, matrices, vectors, and scalars are denoted by letters of bold upper case, bold lower case, and lower case, respectively. Signs $(\bullet)^\ast$,  $(\bullet)^T$, and Tr$(\bullet)$ denote matrix conjugate, transpose, and trace, respectively. 

\section{System model}
For a multi-user  MIMO system with single BS and $K$ users, the MF beamforming vector of user $k$ at BS is defined as ${\bf{v}}_k{\rm{=}}\frac{{\bf{h}}_k^\ast}{\parallel{\bf{h}_k}\parallel_2}$, where   each user is assumed to have only one antenna, and ${\bf{h}}_k$ is the $N\times 1$ channel vector of user $k$ with each component being independent ${\mathcal{CN}}(0,\sigma_h^2)$ and $N$ being the number of transmit antennas at BS.  When the original MF precoder ${\bf{v}}_k{\rm{=}}\frac{{\bf{h}}_k^\ast}{\parallel{\bf{h}_k}\parallel_2}$ is used at BS, the signal received by user $k$ is
\begin{align}
\label{eq_sys_mod}
r_k {\rm{=}} \rho \frac{1}{\sqrt{N}\sigma_h} {\bf{h}}_k^T \frac{{\bf{h}}_k^\ast}{\parallel{\bf{h}_k}\parallel_2} s_k{\rm{+}} \rho\frac{1}{\sqrt{N}\sigma_h} {\bf{h}}_k^T \sum\limits_{l=1, l\ne k}^K \frac{{\bf{h}}_l^\ast}{\parallel{\bf{h}_l}\parallel_2} s_l {\rm{+}}n_k,
\end{align}
where $\rho^2$ is  a parameter proportional to the  transmit power at BS, $s_k$ and $s_l$ are the transmitted data symbols for users $k$ and $l$, $n_k$ is the additive white Gaussian noise with mean zero and variance $\sigma_n^2$, and the normalized factor $1/(\sqrt{N}\sigma_h)$ is to ensure that the average received signal-to-noise-ratio (SNR) is equal to $\rho^2/\sigma^2_n$. From (\ref{eq_sys_mod}), the SINR for user $k$ is given by
\begin{align}
\label{eq_real_SINR}
\gamma_k=\frac{\frac{\rho^2}{N\sigma^2_h}\left(\mathbf{h}_k^T\mathbf{h}_k^\ast\right)}
{\sigma_n^2+ \frac{\rho^2}{N\sigma_h^2} \sum\limits_{l=1, l\ne k}^K \mathbf{h}_k^T\frac{\mathbf{h}_l^\ast}{\parallel{\mathbf{h}_l}\parallel_2}\frac{\mathbf{h}_l^T}{\parallel{\mathbf{h}_l}\parallel_2}\mathbf{h}_k^\ast}.
\end{align}
which  can be written as
\begin{align}
\begin{split}
\label{eq_appr_SINR}
\gamma_k&=\frac{\frac{\rho^2}{N\sigma^2_h}\left(\mathbf{h}_k^T\mathbf{h}_k^\ast\right)}
{\sigma_n^2+ \frac{\rho^2}{N\sigma_h^2}{\parallel{\mathbf{h}_k}\parallel_2^2} \sum\limits_{l=1, l\ne k}^K \frac{\mathbf{h}_k^T}{\parallel{\mathbf{h}_k}\parallel_2}\frac{\mathbf{h}_l^\ast}{\parallel{\mathbf{h}_l}\parallel_2}\frac{\mathbf{h}_l^T}{\parallel{\mathbf{h}_l}\parallel_2}\frac{\mathbf{h}_k^\ast}{\parallel{\mathbf{h}_k}\parallel_2}} \\
&=\frac{\frac{\rho^2}{N\sigma^2_h}\left(\mathbf{h}_k^T\mathbf{h}_k^\ast\right)}
{\sigma_n^2+ \frac{\rho^2}{N\sigma^2_h}\left(\mathbf{h}_k^T\mathbf{h}_k^\ast\right) \sum\limits_{l=1, l\ne k}^K \mathbf{h}_k'^T\mathbf{h}_l'^\ast\mathbf{h}_l'^T\mathbf{h}_k'^\ast}.
\end{split}
\end{align}
where
\begin{align}
\mathbf{h}_k'=\frac{\mathbf{h}_k}{\parallel{\mathbf{h}_k}\parallel_2},
\end{align}
and
\begin{align}
\mathbf{h}_l'=\frac{\mathbf{h}_l}{\parallel{\mathbf{h}_l}\parallel_2}
\end{align}
The sum-rate of all users is
\begin{align}
\label{eq_sumrate}
R_{sum}=\sum^K_{k=1}\log_2\left(1+\gamma_k\right).
\end{align}

\section{Derived Approximate PDF of SINR}
For convenience of deriving the PDF of the SINR defined in (\ref{eq_appr_SINR}), we first define
\begin{align}
\label{eq_sig_power}
x=\frac{\rho^2}{N\sigma^2_h}\mathbf{h}_k^T\mathbf{h}_k^\ast,
\end{align}
and
\begin{align}
\label{eq_Int_l_term}
y_{l}=\mathbf{h}_{k'}^T\mathbf{h}_{l'}^\ast,
\end{align}
Without loss of generality, we can rewrite (\ref{eq_appr_SINR}) as
\begin{align}
\label{eq_simple_sinr}
\gamma=\frac{x}{\sigma_n^2+x\sum\limits_{l=1, l\ne k}^Ky_ly_l^\ast}.
\end{align}
From (\ref{eq_simple_sinr}), it is obvious that we can derive the probability density function (PDF) of $\gamma$ provided that the PDFs of $x$ and $y$ are known. Therefore, our first task is to calculate the PDF formulas of random variables $x$ and $y$.

Random variable $x$ in (\ref{eq_sig_power}) is expanded as a sum of squares of  $2N$ independent and identically distributed real Gaussian random variables
\begin{align}
\label{eq4_sig_expand}
x{\rm{=}}\frac{\rho^2}{2N}\underbrace{\sum\limits_{i=1}^{N} \left(\left[ \frac{\sqrt{2}}{\sigma_h}{\rm{Re}}\left(h_{ki}\right) \right]^2 {\rm{+}} \left[ \frac{\sqrt{2}}{\sigma_h}{\rm{Im}} \left(h_{ki}\right) \right]^2 \right)}_{\tilde{x}},
\end{align}
where ${\rm{Re}}(a)$ and ${\rm{Im}}(a)$ represent the real and imaginary parts of complex scalar $a$, respectively. Obviously,  the random variable $\tilde{x}$ in (\ref{eq4_sig_expand}) is a sum of squares of  $2N$ independent and identically distributed real Gaussian random variables with zero mean and unit variance,  and has a central $\chi ^2(2N)$ distribution which yields the PDF of $x$ from (\ref{eq4_sig_expand})
\begin{align}
\label{eq_x_pdf}
f_x(x)= \frac{x^{N-1}e^{-\frac{x}{2c_S}}}{(2C_S)^N\Gamma(N)},
\end{align}
with
\begin{align}
C_S =\frac{\rho^2}{2N},
\end{align}
and
\begin{align}
\Gamma(a) =\int_0^{+\infty}t^{a-1}e^{-t}dt.
\end{align}
From the definition of $y_{l}$ in (\ref{eq_Int_l_term}), $y_{l}$ is expanded as
\begin{align}
\label{eq_Int_l_term_exp}
y_l=\sum\limits_{n=1}^N h_{kn}'h_{ln}'^\ast,
\end{align}
which is regarded as the sum of $N$ independent random variables  $h_{kn}'h_{ln}'^\ast$ for $n\in \left\{1,2,\cdots, N\right\}$. Due to the central limit theorem, $y_l$ can be approximated as a complex Gaussian distribution with mean $\mu_y$  and variance $\sigma_y^2$ for sufficiently large $N$.
Since $h_{km}'$ and $h_{lm}'$ are independent of each other for all $k\neq l$ or $n\neq m$,  their products are also independent. Therefore,we may find their means as
\begin{align}
\label{eq_iid_zero}
E\left(h_{km}'h_{lm}'^\ast \right)=0.
\end{align}
for all cases of $k\neq l$ or $n\neq m$.
Then, the expected value of $y$ is given by
\begin{align}
\mu_{y}=E(y_l)=0,
\end{align}
and its variance is
\begin{align}
\label{eq_inte_var}
\sigma_{y}^2=\text{E}\left\{ [y_l-E(y_l)]^2 \right\}=\text{E}\left(y_l^2\right)-
\left[\mu_{y_l}\right]^2=\sum\limits_{m=1}^N\sum\limits_{n=1}^N\text{E}\left(h_{km}'h_{lm}'^\ast h_{ln}'h_{kn}'^\ast \right),
\end{align}
where the expected value of the product of four complex gains within sum operation is equal to $\frac{1}{N^2}$  except for  $m=n$,  and zero for other situations due to the independent property between these random variables with zero mean. Thus, $\sigma_y^2$   is given by
\begin{align}
\sigma_y^2=\frac{1}{N}.
\end{align}

Let us define
\begin{align}
\label{eq_inte_power}
z=\sum\limits_{l=1, l\ne k}^Ky_ly_l^\ast=C_I\underbrace{\sum\limits_{l=1, l\ne k}^K\left[\text{Re}\left({\tilde{y}}_l\right)\right]^2+
\left[\text{Im}\left({\tilde{y}}_l\right)\right]^2}_{\tilde{z}},
\end{align}
where $ \text{Re}\left({\tilde{y}}_{l}\right)$ and $\text{Im}\left({\tilde{y}}_{l}\right)$ are the normalized real and image parts of $y_{l}$ with zero mean and unit variance, respectively,
\begin{align}
C_I=\frac{\sigma_y^2}{2}=\frac{1}{2N}.
\end{align}
From the chi-squared distribution definition, the random variable  $\tilde{z}$ in (\ref{eq_inte_power}) obeys a central chi-squared distribution with $2K-2$ degrees of freedom. Thus,  the random variable $z$ has the PDF as follows
\begin{align}
f_{z}(z)= \frac{z^{K-2}e^{-\frac{z}{2C_I}}}{(2C_I)^{K-1}\Gamma(K-1)},
\end{align}
Considering random variables $x$ and $z$ are independent with each other, the joint probability of independent variables  $x$ and $z$ is
\begin{align}
\label{eq_joint_prob}
f_{xz}(x,z)=f_x(x)f_z(z).
\end{align}

In accordance with the definitions of $\gamma$,  $x$ and $z$, we have the relationship among them
\begin{align}
\gamma=\frac{x}{\sigma_n^2+xz}=\frac{1}{t+z},
\end{align}
with
\begin{align}
\label{eq_t_exp}
t=\frac{\sigma_n^2}{x}.
\end{align}
Based on (\ref{eq_t_exp}),  using the Jacobian transformation,  we can derive the PDF of $t$  as follows
\begin{align}
f_t(t)=\frac{\sigma_n^{2N}}{t^{N+1}}\frac{e^{-\frac{\sigma_n^2}{2C_St}}}{(2C_S)^{N}\Gamma(N)}.
\end{align}
Let us define $w=t+z$, then we have the PDF of random variable $w$
\begin{align}
\begin{split}
f_w(w)=\int^{w}_{0}f_{z}(z)f_{t}(w-z)dz=\int^{w}_{0}\frac{\sigma_n^{2N}e^{\frac{\sigma_n^2}{2C_S(z-w)}}z^{K-2}e^{\frac{-z}{2C_I}}}{(w-z)^{N+1}(2C_S)^{N}\Gamma(N)(2C_I)^{K-1}\Gamma(K-1)}dz.
\end{split}
\end{align}
Considering
\begin{align}
\gamma=\frac{1}{w},
\end{align}
using the single-variable Jacobian transformation, we have the resulting PDF of the SINR $\gamma$
\begin{align}
\label{eq_gamma}
f_{\gamma}(\gamma)&=\frac{1}{\gamma^2}f_w(\frac{1}{\gamma})=\frac{1}{\gamma^2}\int^{\frac{1}{\gamma}}_{0}\frac{\sigma_n^{2N}e^{\frac{\sigma_n^2}{2C_S(z-\frac{1}{\gamma})}}z^{K-2}e^{\frac{-z}{2C_I}}}{(\frac{1}{\gamma}-z)^{N+1}(2C_S)^{N}\Gamma(N)(2C_I)^{K-1}\Gamma(K-1)}dz. \end{align}
If we define $\nu=(\frac{1}{\gamma}-z)$, the above integral can be further simplified  as
\begin{align}
\label{eq_gamma_pdf}
f_{\gamma}(\gamma)=C_\gamma\int^{\frac{1}{\gamma}}_{0}\frac{\left(\gamma^{-1}-\nu\right)^{K-2}e^{N\nu-\frac{N\sigma_n^2}{\rho^2}\nu^{-1}}}{\nu^{N+1}}d\nu,
\end{align}
where
\begin{align}
C_\gamma=\frac{\sigma_n^{2N}e^{-\frac{1}{2C_I\gamma}}}{(2C_S)^{N}\Gamma(N)(2C_I)^{K-1}\Gamma(K-1)\gamma^2}=\frac{\sigma_n^{2N}e^{-\frac{N}{\gamma}}N^{K+N-1}}{(\rho)^{2N}\Gamma(N)\Gamma(K-1)\gamma^2}.
\end{align}
This completes the derivation of the SINR PDF of user $k$.

Using (\ref{eq_gamma_pdf}), we readily attain the following average symbol error rate
\begin{align}
\bar{P}_{SER}=\tilde{\alpha}\int_0^{+\infty}Q\left(\tilde{\beta}\gamma\right) f_{\gamma}(\gamma)d\gamma,
\end{align}
and average sum-rate
\begin{align}
\bar{R}_{sum}=K\int_0^{+\infty}\log_2\left(1+\gamma\right) f_{\gamma}(\gamma)d\gamma,
\end{align}
where $\tilde{\alpha}$ and $\tilde{\beta}$ depend on the  modulation method in \cite{Andr}, and the $Q$ function is specified below
\begin{align}
Q(x)=\int_x^{+\infty}\frac{1}{\sqrt{2\pi}}\exp{\left(-\frac{t^2}{2}\right)}.
\end{align}
\section{SIMULATION AND DISCUSSION}
In order to evaluate the validity of the derived approximate PDF of SINR corresponding to the beamformer MF, by randomly generating many samples of $\bf{h}_k$, we calculate the exact values of SINR by using (\ref{eq_real_SINR}) and adopt the function ksdensity() in Matlab software to output the simulated PDF provided that parameters $\rho$, $N$, $K$, $\sigma^2_h$, and $\sigma^2_n$ are fixed. The simulated exact PDF is used as a reference and compared with the derived approximate PDF of SINR directly computed by (\ref{eq_gamma_pdf}).

\begin{figure}[H]
\centering
\includegraphics[scale=0.5]
 {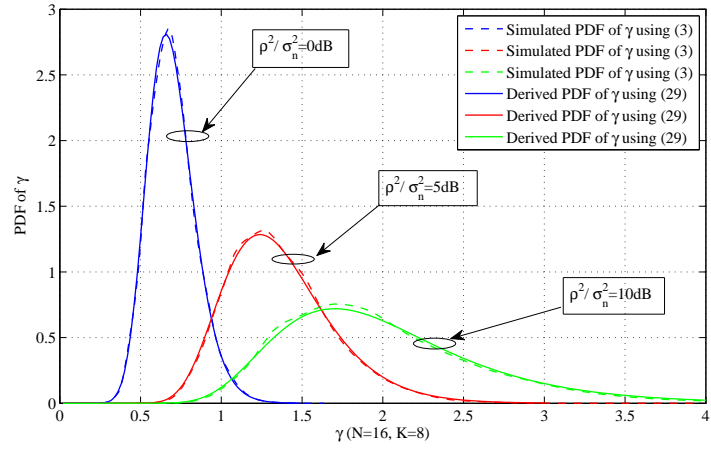}\caption{PDF of SINR $(N=16,K=8)$.}\label{Fig1}
\end{figure}
\begin{figure}[H]
\centering
\includegraphics[scale=0.5]
 {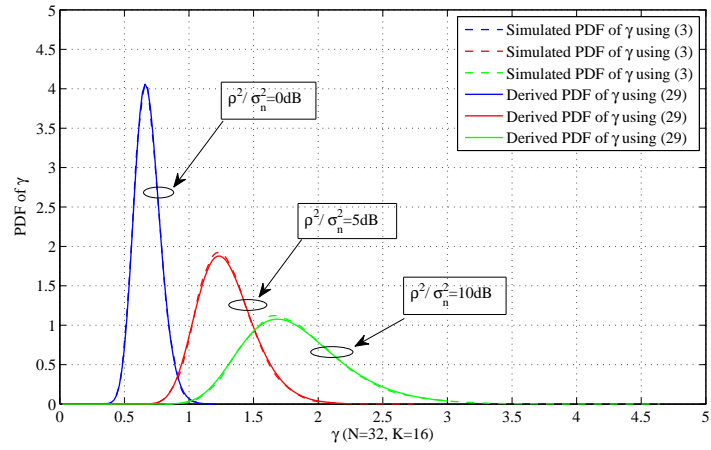}\caption{PDF of SINR $(N=32,K=16)$.}\label{Fig2}
\end{figure}

\begin{figure}[H]
\centering
\includegraphics[scale=0.5]
 {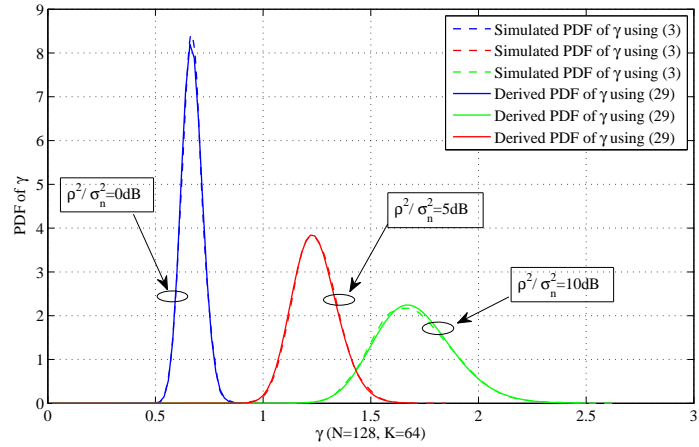}\caption{PDF of SINR $(N=128,K=64)$.}\label{Fig4}
\end{figure}
Fig.~1, Fig.~2, and Fig.~3 demonstrates the curves of PDF of SINR for different typical SNRs defined as  $\rho^2/\sigma^2_n$ for $N=2K$. Obviously, the derived PDFs are in very good agreement with the simulated PDFs for three different scenarios: $N=16$, $N=32$, and $N=128$.
\begin{figure}[H]
\centering
\includegraphics[scale=0.5]
 {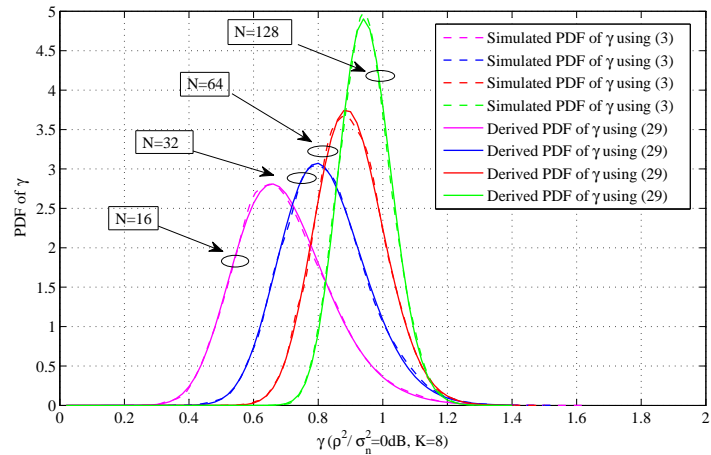}\caption{PDF of SINR $(\rho^2/ \sigma^2_n=0dB,K=8)$.}\label{Fig5}
\end{figure}\begin{figure}[H]
\centering
\includegraphics[scale=0.5]
 {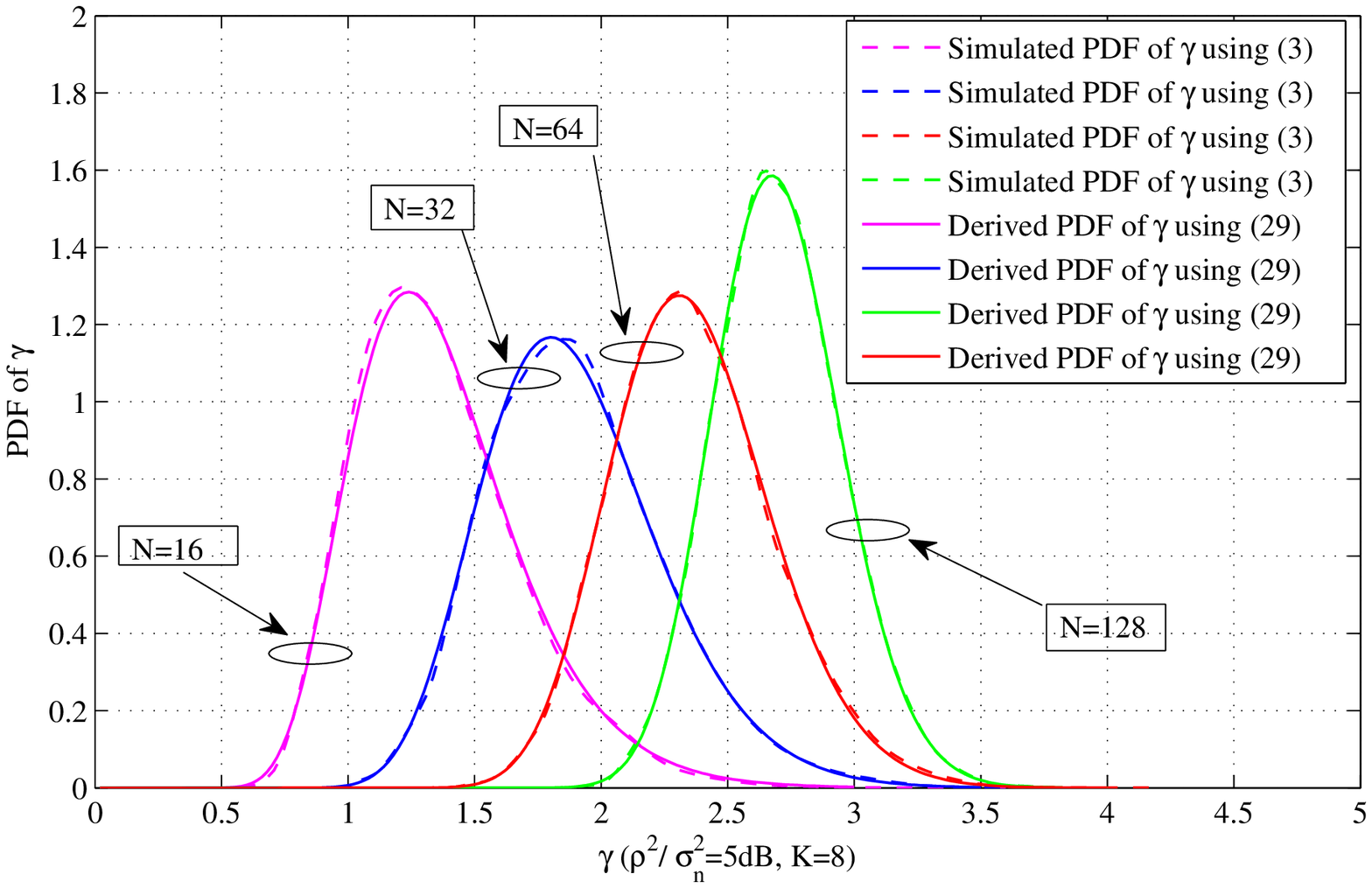}\caption{PDF of SINR $(\rho^2/ \sigma^2_n=5dB,K=8)$.}\label{Fig6}
\end{figure}\begin{figure}[H]
\centering
\includegraphics[scale=0.5]
 {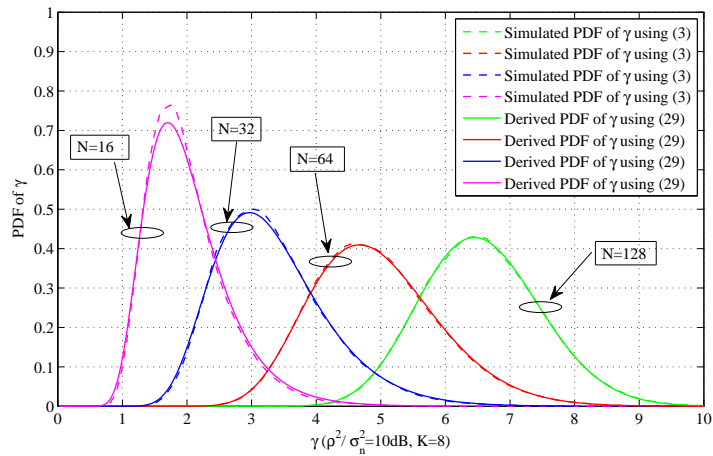}\caption{PDF of SINR $(\rho^2/ \sigma^2_n=10dB,K=8)$.}\label{Fig7}
\end{figure}
Fig.~4, Fig.~5, and Fig.~6 shows the curves of PDF of SINR for different typical values of $N$  for fixed  $K=8$ and SNR. Obviously, the derived PDF matches well with the simulated PDF for three different scenarios: $\rho^2/ \sigma^2_n=0dB$, $\rho^2/ \sigma^2_n=5dB$ and $\rho^2/ \sigma^2_n=10dB$.

From the above six figures, we also find,  as $N$ increases, the simulated and derived PDFs of SINR converge to the same Gaussian distribution.

\section{Conclusion}
We derived an approximate expression for the probability density function of SINR for medium-scale and large-scale MU-MIMO systems with MF beamformer. It is confirmed that the derived probability density function is very close to its counterpart obtained by simulation. The procedure of using the derived function to calculate the probability of symbol error is presented, which can also be applied to analyze the outage probability and average capacity.

\appendices


\ifCLASSOPTIONcaptionsoff
  \newpage
\fi

%
%
%
%
\newpage

\end{document}